\documentclass[11pt,a4paper]{article}
\pdfoutput=1
\usepackage{jinstpub}

\usepackage{amsmath}
\usepackage{graphicx}
\usepackage{hyperref}
\usepackage{array}
\usepackage{subcaption}
\usepackage[usenames,dvipsnames]{xcolor}

\usepackage{lineno}

\title{Suppression of accidental backgrounds with deep neural networks in the PandaX-II experiment}
\author[a]{Nasir Shaheed}
\author[b,c,1]{Xun Chen,\note{Corresponding author.}}
\author[a,2]{Meng Wang,\note{Corresponding author.}}

\affiliation[a]{Research Center for Particle Science and Technology, Institute of Frontier and Interdisciplinary Science, Shandong University, Qingdao 266237, Shandong, China}
\affiliation[b]{INPAC and School of Physics and Astronomy, Shanghai Jiao Tong University, MOE Key Lab for Particle Physics, Astrophysics and Cosmology, Shanghai Key Laboratory for Particle Physics and Cosmology,\\ Shanghai 200240, China}
\affiliation[c]{Shanghai Jiao Tong University Sichuan Research Institute,\\ Chengdu 610213, China}

\emailAdd{chenxun@sjtu.edu.cn}
\emailAdd{mwang@sdu.edu.cn}

\abstract{ The PandaX dark matter detection project searches for dark
  matter particles using the technology of dual phase xenon time
  projection chamber. The low expected rate of the signal events makes
  the control of backgrounds crucial for the experiment success. In
  addition to reducing external and internal backgrounds during the
  construction and operation of the detector, special techniques are
  employed to suppress the background events during the data analysis.
  In this article, we demonstrate the use of deep neural networks
  (DNNs) for suppressing the accidental backgrounds, as an alternative
  to the boosted-decision-tree method used in previous analysis of
  PandaX-II. A new data preparation approach is proposed to enhance
  the stability of the machine learning algorithms to be run and
  ultimately the sensitivity of the final data analysis.  }

\begin{document}

\maketitle
\flushbottom

\section{Introduction}
\label{sec:intro}
The nature of dark matter in our universe remains one of the most
fundamental unresolved questions in physics. The weakly interacting
massive particles (WIMPs) proposed by various theories beyond the
Standard Model of particle physics, are considered as a leading candidate for
dark matter~\cite{Bertone:2004pz}. In recent decades, numerous
experiments have been conducted to detect direct collisions between
WIMPs and ordinary matter in deep underground laboratories, resulting
in the suppression of the available parameter space for
WIMPs~\cite{Liu:2017drf,Billard:2021uyg}. Among these projects are the
PandaX-II~\cite{PandaX:2016pdl} and PandaX-4T~\cite{PandaX:2018wtu}
experiments, located at the China Jinping Laboratory
(CJPL)~\cite{Wu:2013cno,Li:2014rca,Cheng:2017usi}, which utilize the
technology of a dual-phase xenon time projection chamber
(TPC)~\cite{Aprile:2009dv}. Recently, the PandaX-4T experiment has
established a more stringent constraint on the spin-independent
interactions between WIMPs and nucleons than previous generations of
the same type of experiments~\cite{PandaX-4T:2021bab}.

Background control is a crucial aspect of experiments searching for
dark matter due to the extremely low rate of the signal particles.  The
main sources of background are gamma or neutron collisions inside the
TPC, which originate from known radioactive sources in the detector
material or dissolved sources within the xenon. The PandaX
collaboration has made significant efforts to reduce backgrounds from
laboratory, detector materials, and the xenon recirculating
pipelines~\cite{Zhang:2016pgh,PandaX-4T:2021lbm,Cui:2020bwf,Wang:2022hkk}. Nevertheless,
accidental background still plays a significant role and must be taken
into account during data analysis.

The PandaX-II experiment successfully utilized the
boosted-decision-tree (BDT) method, a machine learning technique, to
suppress accidental background~\cite{PandaX-II:2022waa}. With the
advancement of deep learning technologies, such as deep neural
networks (DNNs), these have become valuable tools in various studies of
particle physics in recent years~\cite{Guest:2018yhq, Carleo:2019ptp,
  Schwartz:2021ftp}. In particular, in the field of deep underground
experiments, DNNs have been widely used to discriminate signals from
backgrounds~\cite{Qiao:2018edn, NEXT:2020jmz, Khosa:2019qgp},
reconstruct the energy and position of events~\cite{EXO:2018bpx,
  Domine:2019zhm}, and improve the speed of data
fitting~\cite{LUX:2022vee}.

In this article, we conduct a study on using DNNs to suppress
accidental background in the PandaX-II experiment. In
Section~\ref{sec:pandax_acc}, we provide a brief overview of the
detection principle and accidental backgrounds in PandaX-II. We then
discuss data preparation in Section~\ref{sec:data}. In
Section~\ref{sec:dnn}, we present the testing of various DNN
architectures for accidental background suppression, as well as a new
data preparation method to improve stability. Finally, we summarize
our findings and provide an outlook in Section~\ref{sec:summary}.

\section{Accidental background in PandaX-II}
\label{sec:pandax_acc}
The central components of the PandaX-II and PandaX-4T detectors are
the dual phase xenon TPCs. They have a similar structure. A TPC has a
cylindrical sensitive volume enclosed by polytetrafluoroethylene
(PTFE) reflection panels. A cathode mesh at the TPC's base and a gate
grid electrode beneath the liquid xenon surface create the drift
field. The gate, in conjunction with the anode mesh above the liquid
level, generates an extraction field which extracts electrons from the
liquid xenon into the gas xenon. Two arrays of photomultiplier tubes
(PMTs) are placed above and below the TPC to detect the scintillation
photons generated within the TPC. For more detailed information on the
PandaX detectors, refer to references ~\cite{PandaX:2016pdl} and
~\cite{PandaX-4T:2021bab}.

To aid in understanding the origin of accidental background, we
provide a brief overview of the detection principle of the dual phase
TPC in this article. For a more in-depth explanation, we refer the
readers to Ref.~\cite{Aprile:2009dv}. The collisions between incoming
particles and target xenon atoms in the dual phase xenon TPC may
produce prompt scintillation photons ($S1$) and ionized electrons. The
electrons drift along the drifting field in the TPC and are extracted
into the gaseous region, where they produce delayed photons ($S2$)
through the process of electroluminescence. The time difference
between the $S1$ and $S2$ signals can be used to determine the
$z$-position of the collision. Additionally, the ratio of $S2/S1$ is
an important discriminator between electron recoil (ER) and nuclear
recoil (NR) events. NR events, which are the interactions of interest
for detecting WIMPs and the neutron backgrounds, are characterized by
a lower ratio of $S2/S1$ compared to ER events due to the fact that a
large fraction of recoil energy converts into heat and escapes
detection. This allows for the discrimination of WIMPs from most of
the background events.

In the PandaX-II and PandaX-4T experiments, the majority of
backgrounds are caused by scattering events of gamma or neutron
originating from radioactive isotopes in the detector materials and
dissolved radioactive isotopes in liquid xenon, such as $^{222}$Rn,
$^{85}$Kr, or $^{3}$H. These backgrounds are controlled through
techniques such as material screening and selection, and xenon
distillation and purification. Another type of background, known as
surface background, is generated by the $\beta$-decay of $^{210}$Pb on
the inner surface of the TPC, which affects the $S2$ signal and is
concentrated near the edges. This background can be modeled using a
data-driven method and estimated.

Accidental background refers to events where $S1$ and $S2$ signals are
not from the same collision events. Identifying and controlling these
backgrounds can be challenging, but they are crucial for achieving a
robust understanding of the signals in the detector.  The $S1$ or
$S2$-like signals are not correlated with any other recorded signals
from the same source are referred to be ``isolated''. During event
reconstruction, unrelated isolated $S1$ and $S2$ signals may appear in
the same drift window, resulting in accidental backgrounds. In
Ref.~\cite{PandaX-II:2022waa}, the possible origins of isolated
signals are analyzed in detail. We present a brief overview of them
here. The isolated $S1$ signals may originate from regular scattering
events, but the corresponding $S2$ signals are not produced or
recorded. They could also be single electron signals that were
misidentified as $S1$s.  Additionally, overlapped dark noises from
different photo-multiplier tubes (PMTs) may form $S1$-like signals.
The isolated $S2$ signals are produced by the electroluminescent
process of electrons in the gas region, similar to regular $S2$
signals. They can be regular $S2$ signals without corresponding $S1$
signals recorded, or overlap with $S1$ signals in such a way that only
the $S2$ signals are recognized. Additionally, stagnant electrons
created by large energy depositions may be randomly released into the
gas region, resulting in $S2$-like signals directly.

\begin{figure}[hbt]
  \centering
  \includegraphics[width=0.6\linewidth]{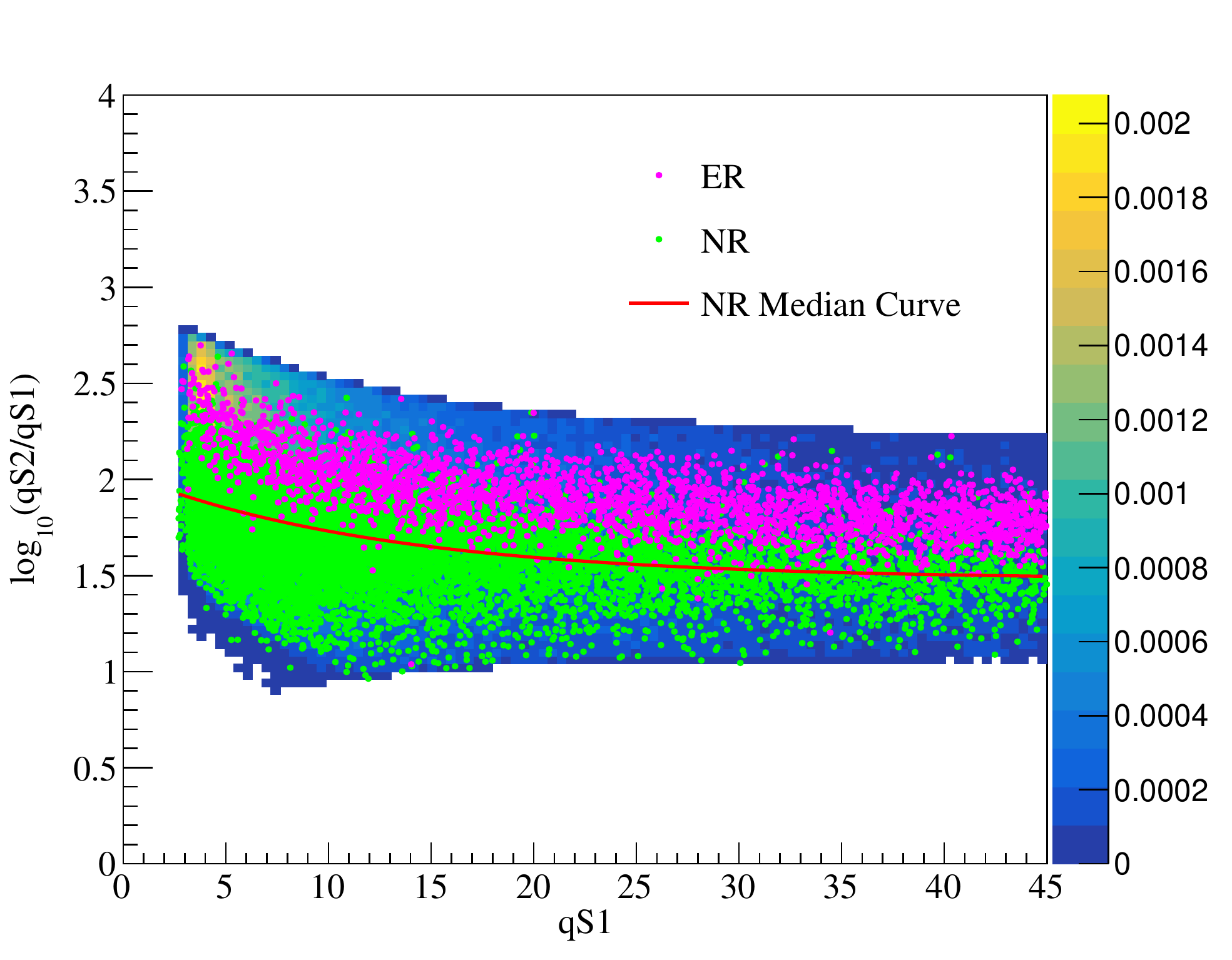}
  \caption{The signal distributions of NR and ER calibration events
    (scatter points) on top of the expected accidental background
    events (background histogram) in Run 11. The reference NR median
    line is plotted.}
  \label{fig:sig_dis_cal}
\end{figure}

Ref.~\cite{PandaX-II:2022waa} estimated the average rates of isolated
$S1$ and $S2$ signals, $\bar{r}_1$ and $\bar{r}_2$, using several
data-driven methods, and obtained consistent results. The total number
of accidental background events was calculated using the equation:
\begin{equation}
  \label{eq:nacc}
  n_{\mbox{acc}} = \bar{r}_1\cdot \bar{r}_2 \cdot \Delta t_{w} \cdot T \cdot \epsilon,
\end{equation}
where $\Delta t_{w}$ is the time window defined by the fiducial volume
cut, $T$ is the duration of the science data run, and $\epsilon$ is
the efficiency of data quality cuts. The final number of accidental
background events in PandaX-II is non-trivial, particularly in the
region beneath the reference median line of NR events in the plot of
$\log (S2/S1)$ versus $S1$ from neutron calibration, where the
statistics for regular ER backgrounds are
low. Fig.~\ref{fig:sig_dis_cal} shows the signal distributions of the
NR and ER calibration events and the accidental background in
PandaX-II Run 11 data set~\cite{Wang:2020coa} for reader's
reference. Backgrounds in this region can obscure or mimic the true
WIMP signals. Suppressing this type of background events is crucial
for the sensitivity of the WIMP search.

In the PandaX-II experiment, the BDT algorithm was employed to
effectively reduce the number of accidental background
events~\cite{PandaX-II:2022waa}. This technique utilizes a set of
characteristics specific to the signals to differentiate between
accidental background and NR events. Given the importance of events
below the NR median line in the search for WIMPs, only samples in that
region were used in the BDT training process. The optimized BDT cuts
achieved a powerful capability to identify accidental background
while maintaining high efficiency for NR and ER signal events. For
instance, the number of expected accidental events below the NR median
in PandaX-II Run 11 was reduced from 2.93 to 0.77, while the
efficiency for NR signal events in the same region is 90.7\%. The
successful implementation of the BDT technique in PandaX-II motivated
us to investigate the use of advanced deep learning techniques for
suppressing accidental background.

\section{Data preparation}
\label{sec:data}

In order to implement DNNs for the purpose of suppressing accidental
background in the PandaX-II experiment, the data preparation is
crucial. The data samples used in this study are sourced from the
PandaX-II Run 11 dataset~\cite{Wang:2020coa}, which were also used in
the previous study with BDT. This allows for consistency in the input
variables and allows for a direct comparison of the performance of the
two methods.

The PandaX experiments take data by digitizing the output voltage of
the PMTs into waveforms. In PandaX-II, the digitized samples within
the $[-500, 500]$~$\mu$s window of a global trigger are
kept~\cite{Wu:2017cjl}. The raw data are processed following the same
procedure: 1) the region over a given threshold in each recorded
waveform segment is marked as ``hit''; 2) hits with tight time
correlation are clustered into ``signal''; 3) for each signal, related
properties are calculated, and the signal is tagged.  The data is
converted to collections of events with signals in the end of data
processing, with properties calculated. The important properties of an
event include:
\begin{itemize}
\item number of S1 signals,
\item number of S2 signals,
\item start time of the event,
\item time difference of each signal to the start of the event,
\item index of the maximum $S1$ before the maximum $S2$,
\item index of the maximum $S2$,
\item summation of extra signals except for the paired maximum $S1$
  and maximum $S2$,
\item energy of the event by combining the paired $S1$ and $S2$.
\end{itemize}

In the search for WIMPs, events featuring a primary $S1$ and $S2$
signal are selected for analysis. The horizontal location of an event
is determined using various techniques based on the top pattern of the
$S2$ signal, while the vertical position is determined by the temporal
separation of the paired $S1$ and $S2$ signals. These signals are then
corrected according to their position. The final data used for
analysis includes key characteristics of each signal, which are
calculated from the waveform and summarized in
Table~\ref{tab:sig_properties}.
\begin{table}[htb]
  \centering
  \begin{tabular}{cp{.6\linewidth}}
    Name & Description \\\hline\hline
    qS1 & raw charge of $S1$, in the unit of photoelectron (PE) \\
    qS2 & raw charge of $S2$ \\
    qS1c & position corrected charge of $S1$ \\
    qS2c & position corrected charge of $S2$ \\
    wS1 & width of the $S1$ \\
    wS2 & width of the $S2$ \\
    wtenS2 & full width of one-tenth maximum of the $S2$ \\
    S1TBA & asymmetry between the charge collected by the top ($q_T$) and bottom ($q_B$) array for $S1$, defined as $(q_T - q_B)/(q_T + q_B)$ \\
    S2TBA & asymmetry between the charge collected by the top and bottom array for $S2$ \\
    S2SY1 & the ratio of charge before the maximum height to the total charge for $S2$, in the raw waveform \\
    S2SY2 & the ratio of charge before the maximum height to the total charge for $S2$, in the smeared waveform \\
    S1NPeaks & number of local maximums in $S1$ \\
    S1LargestBCQ & ratio of the largest charge detected by one bottom PMT to the total charge of $S1$ \\\hline
  \end{tabular}
  \caption{Important properties of the signals used in data analysis.}
  \label{tab:sig_properties}
\end{table}

The aim of this study is to distinguish between accidental background
events below NR median and physical scattering events, therefore, two
types of data samples are necessary.  Samples of accidental
background were generated by randomly pairing isolated $S1$ and $S2$
signals identified in the dark matter search data.  Since the region
below the NR median has sufficient statistics only in the neutron
calibration run, the physical scattering event samples are extracted
from the neutron calibration runs. Additionally, the DNNs are
anticipated to classify the greatest number of physical events above
the NR median correctly, especially the ER events. Therefore, a third
dataset is acquired from the related ER calibration runs.  The events
that have been chosen, which fall within the defined fiducial volume,
have met all the established criteria during the PandaX-II data
analysis process and fall within the signal window, including a charge
of $S1$ within a range of 3 and 45 PE, and a raw charge of $S2$
greater than 100 PE, with a corrected $S2$ charge smaller than 10,000
PE. For the accidental background, only the events below NR median
are selected. The generated data set of accidental background
contains 43,719 events, and the full NR data set contains 10,881
events.

The data samples are structured in the ROOT format, a widely used tool
in high energy physics experiments~\cite{Brun:1997pa}. The variables
are organized as branches within the {\tt TTree} structure, allowing
for easy implementation of a cut to select NR events below the NR
median line during data loading.

In order to train the DNN, 80\% of the input data are used, 10\% of
the data are set aside for validation, and the remaining 10\% for
testing purposes. This split of the data allows for a thorough
evaluation of the performance of the DNN and enables the
identification of any potential issues during the training process.

\section{Deep neural networks}
\label{sec:dnn}
The task of identifying accidental background events can be approached
as a binary classification problem. Since the number of features in
the event data is limited, a Multi-Layer Perceptron
(MLP)~\cite{10.1162/neco.2006.18.7.1527}, a special type of artificial
neural networks, is an appropriate deep learning model for this task.

The basic idea of the artificial neural network is to construct a
function $N$ with many parameters, which maps the input features
$\mathbf{X}$ to the prediction $\mathbf{y}$, or
$\mathbf{y} = N(\mathbf{X})$.  Within the network, there are many
layers of computation units called neurons, which accepts inputs from
neurons from other layers and generate outputs to neurons in the
following layers. By minimizing a given loss function of
$l(\mathbf{y}, \mathbf{Y})$, which takes the prediction $\mathbf{y}$
and the label $\mathbf{Y}$ of data as inputs, the parameters of $N$
are able to be adjusted. This process is generally referred to as
"training" the function $N$. The training process is done by
backpropagation of errors, where the errors are propagated back through
the layers of the network to update the parameters of the function. A
trained function is able to fit the training data well and can be used
to make predictions by feeding new data.

The MLP contains multiple layers of neurons fully connected in a
feedforward manner. The input layer takes in the input features
$\mathbf{X}$, and the output layer produces the prediction
$\mathbf{y}$. In between the input and output layers, there are one or
more hidden layers that help to extract complex features from the
input data. Once the MLP is trained, it can be used to predict by
feeding new data into the input layer.

\begin{figure}[hbt]
  \centering
  \includegraphics[width=0.6\linewidth]{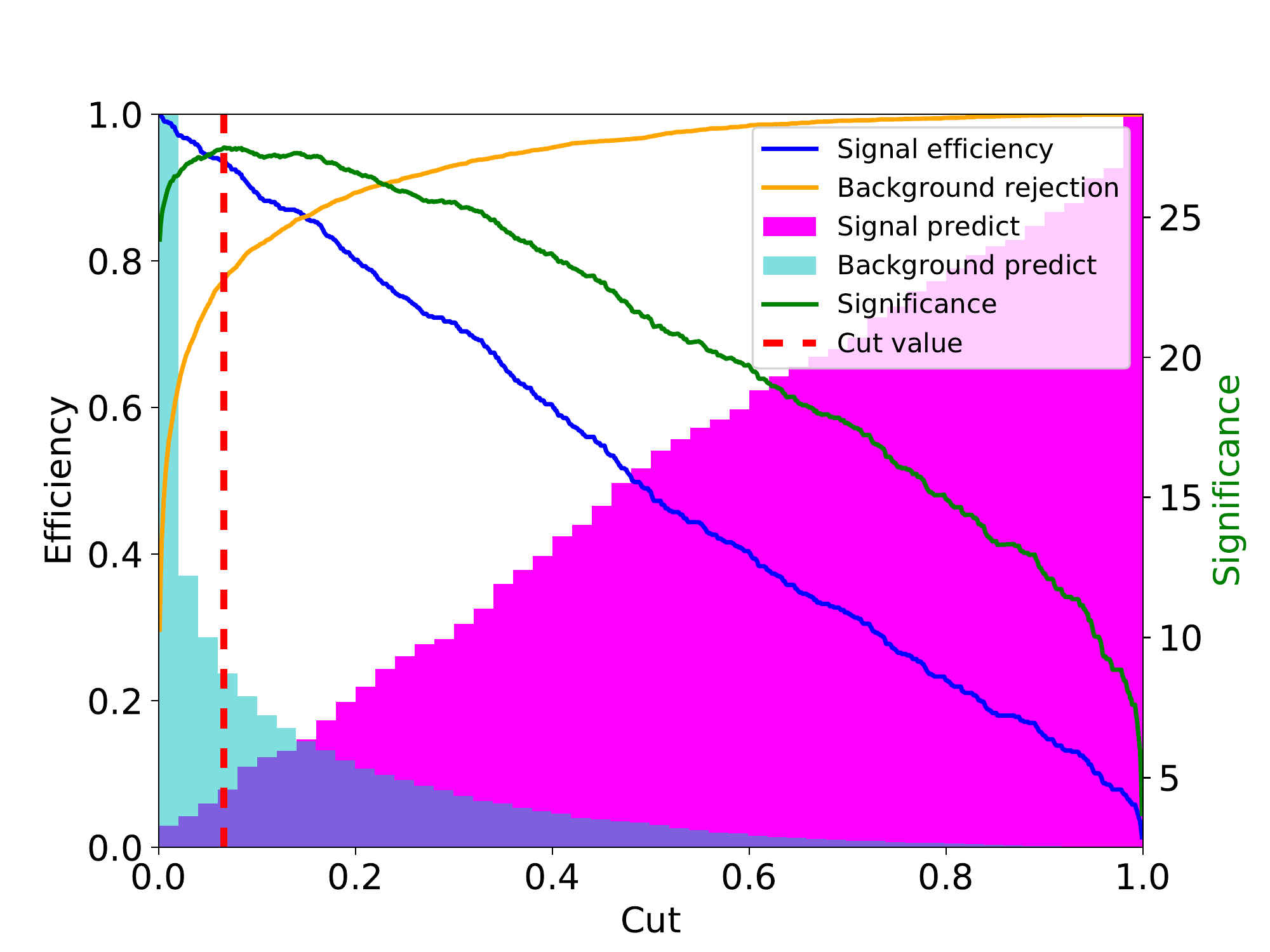}
  \caption{The detemination of the cut value as well as the
    significance $S$. The distributions of the MLP output of input
    test events are plotted as backgrounds. The distributions are
    obtained from one of the best parameters for model 4C (see text
    below).}
  \label{fig:det_significance}
\end{figure}

In this study, we implemented the MLP models with TensorFlow
2.5~\cite{tensorflow2015-whitepaper}. The event properties listed in
Table.~\ref{tab:sig_properties} are used as input features of the
MLP. The input data have been rescaled with the min-max
normalization. The activation function used in the output layer of the
MLP is the ``sigmod'' function, which maps the output to a value
within the range of $[0, 1]$. The value can be utilized to determine
if an event is physical or non-physical according to an optimal cut
value. The cut is obtained by maximizing the significance $S$, the
metric used in the previous study, which is defined as
\begin{equation}
  S = \frac{\epsilon_s n_s}{\sqrt{\epsilon_s n_s + \epsilon_b n_b}},
\label{eq:significance}
\end{equation}
where $n_s$ and $n_b$ are the numbers of NR and accidental background
events, respectively, and the $\epsilon$s are the corresponding
efficiencies obtained with a certain cut value. To conform with the
methodology in the previous study~\cite{PandaX-II:2022waa}, the values
of $n_s$ and $n_b$ are approximated to be equal and used
accordingly.  Fig.~\ref{fig:det_significance} presents an example of
the determination of the significance.

The first objective of this study is to determine the optimal structure of
the hidden layers, including the number of layers and the number of
neurons in each layer.  Thus we conducted a scan across different
layer combinations to find the maximum value of $S$. During the scan,
the learning rate was fixed at 0.001, the loss function was binary
cross entropy, and the optimizer used was
Adam~\cite{2014arXiv1412.6980K}. To make a fair comparison with the
previous study of the BDT method, the training data contains only the
events below the NR median, following the same strategy.

In the training process of each MLP model, a maximum of 1,500 epochs
was set to ensure the model's convergence and to prevent
overfitting. The training stopped when the validation loss function
reached its minimum value with a patience of 20. The network
parameters with the lowest validation loss value were then saved and
used to calculate the significance.
To obtain reliable results, each model was trained from scratch 100
times and the average significance was calculated.

\begin{figure}[hbt]
  \centering
  \includegraphics[width=0.8\linewidth]{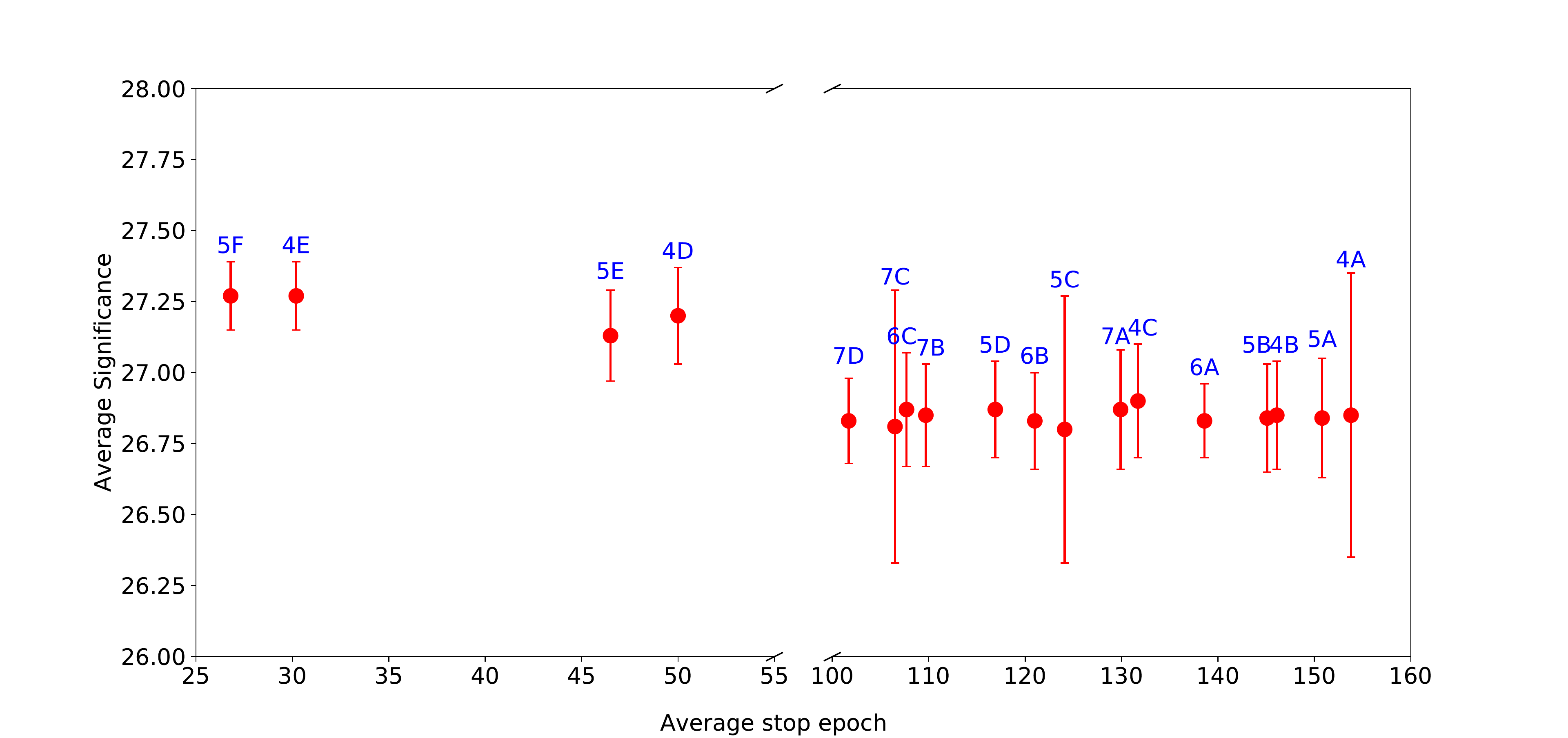}
  \caption{The average significances and epochs of the scanned networks.}
  \label{fig:model_res}
\end{figure}

The results of the tested MLP structures, including the average
significance and the average number of epochs required to reach the
best parameters, are summarized in Table~\ref{tab:model_res} and
visualized in Figure~\ref{fig:model_res}.  It was observed that as the
number of neurons in the network increased, the training process
completed earlier, and the significance is higher.  Additionally, it
is observed that all the deep learning models have achieved a higher
significance compared to the 26.2 value obtained by the BDT method
reported in Ref.~\cite{PandaX-II:2022waa}. The differences in average
significances among the various models are relatively small.

\begin{table}[hbt]
  \centering
  \begin{tabular}{cccc}
    Label & Hidden Layers & Average Significance & Average stop epoch \\\hline
    4A & $32\times16\times8\times8$ & 26.85$\pm$0.50 & 153.8 \\
    4B & $32\times16\times16\times8$ & 26.85$\pm$0.19 & 146.1 \\
    4C & $32\times32\times16\times8$ & 26.90$\pm$0.20 & 131.7 \\
    4D & $128\times128\times64\times32$ & 27.20$\pm$0.17 & 50.0 \\
    4E & $512\times512\times256\times256$ & 27.27$\pm$0.12 & 30.2 \\
    5A & $32\times16\times8\times8\times8$ & 26.84$\pm$0.21 & 150.8 \\
    5B & $32\times16\times16\times8\times8$ & 26.84$\pm$0.19 & 145.1 \\
    5C & $32\times32\times16\times8\times8$ & 26.80$\pm$0.47 & 124.1 \\
    5D & $32\times32\times16\times16\times8$ & 26.87$\pm$0.17 & 116.9 \\
    5E & $128\times128\times64\times32\times16$ & 27.13$\pm$0.16 & 46.5 \\
    5F & $512\times512\times256\times256\times128$ & 27.27$\pm$0.12 & 26.8 \\
    6A & $32\times16\times16\times8\times8\times8$ & 26.83$\pm$0.13 & 138.6 \\
    6B & $32\times32\times16\times16\times8\times8$ & 26.83$\pm$0.17 & 121.0 \\
    6C & $32\times32\times16\times16\times16\times8$ & 26.87$\pm$0.20 & 107.7 \\
    7A & $32\times16\times16\times16\times12\times12\times12$ & 26.87$\pm$0.21 & 129.9 \\
    7B & $32\times32\times16\times16\times16\times8\times8$ & 26.85$\pm$0.18 & 109.7\\
    7C & $32\times32\times16\times16\times12\times12\times8$ & 26.81$\pm$0.48 & 106.5 \\
    7D & $32\times32\times16\times16\times16\times12\times12$ & 26.83$\pm$0.15 & 101.7 \\\hline
  \end{tabular}
  \caption{The average significances and epochs of the scanned
    networks. The numbers in the column ``Hidden Layers'' are
    neurons.}
  \label{tab:model_res}
\end{table}

\begin{figure}[hbt]
  \centering
  \begin{subfigure}[t]{0.4\textwidth}
    \includegraphics[width=\textwidth]{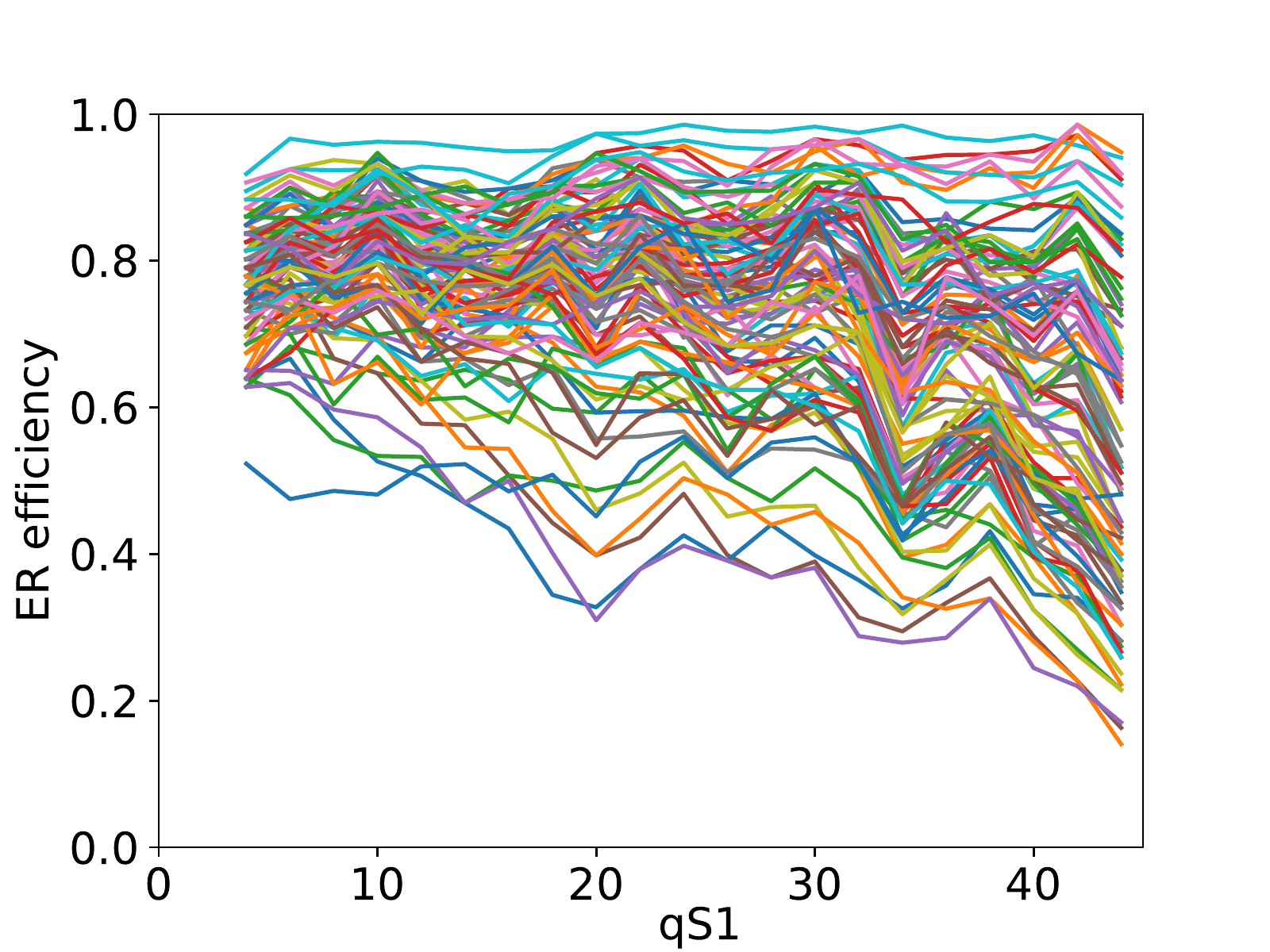}
    \caption{MLP 4C}
  \end{subfigure}
  \begin{subfigure}[t]{0.4\textwidth}
    \includegraphics[width=\textwidth]{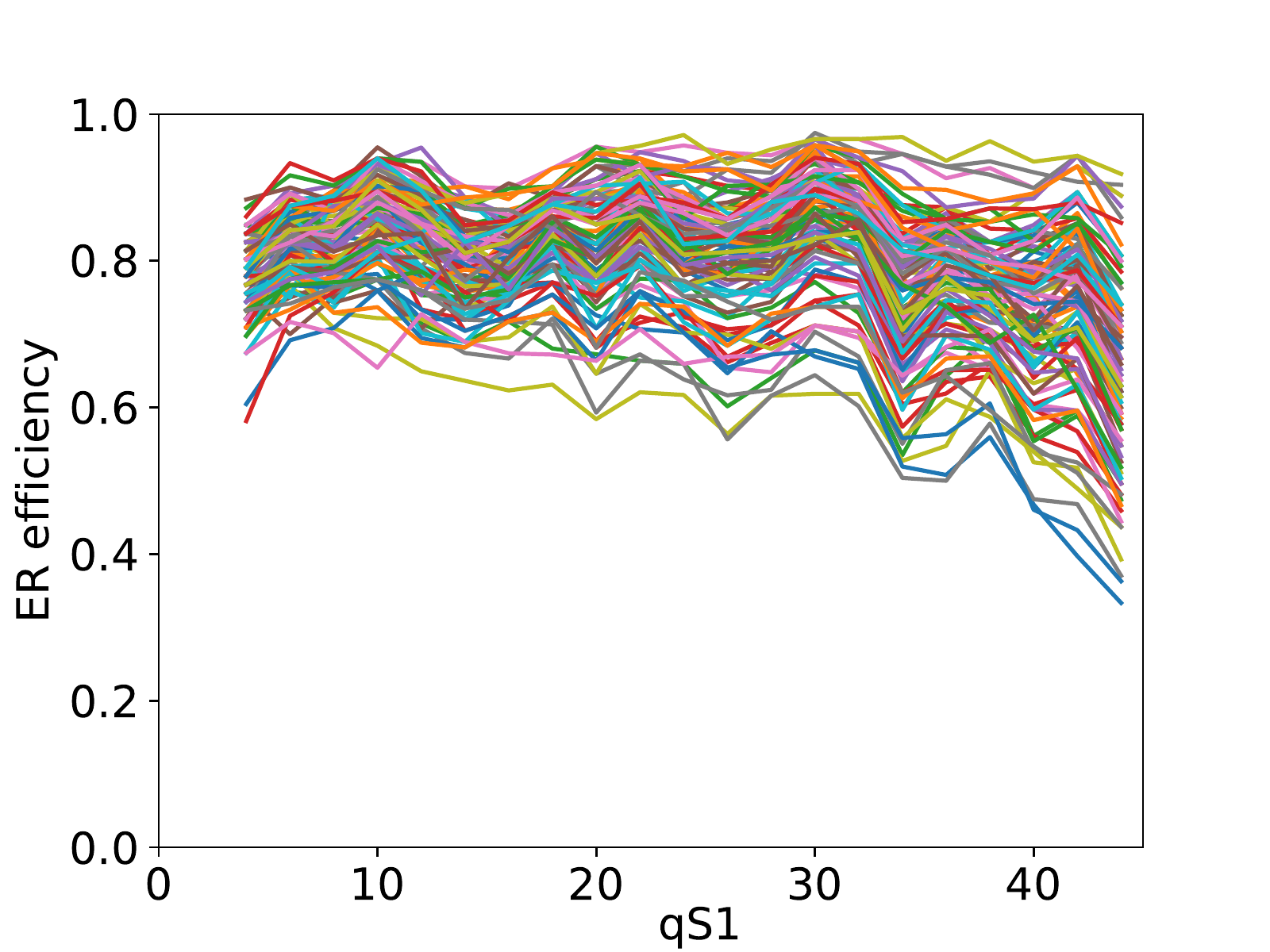}
    \caption{MLP 5E}
  \end{subfigure}
  \begin{subfigure}[t]{0.4\textwidth}
    \includegraphics[width=\textwidth]{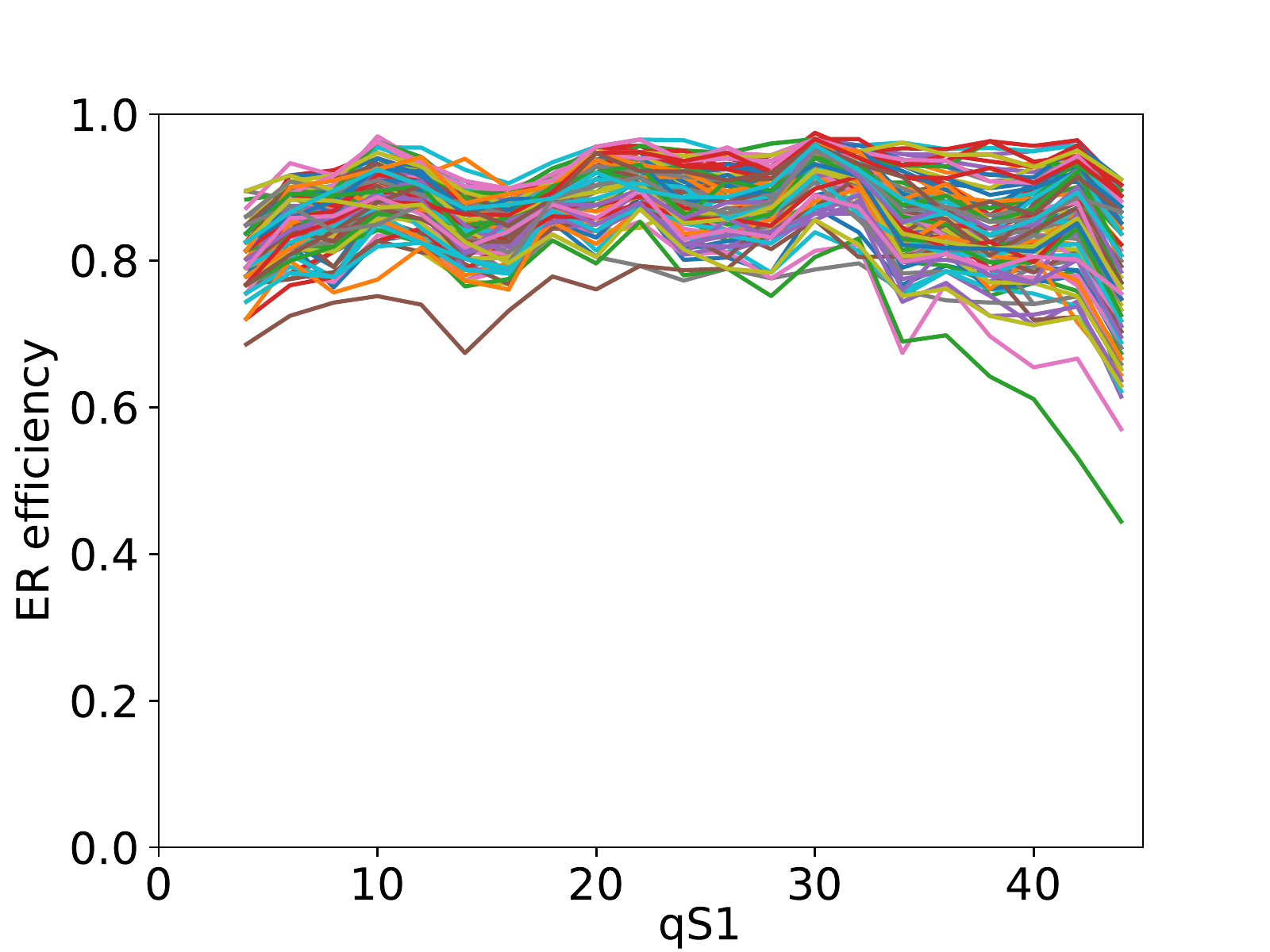}
    \caption{MLP 5F}
  \end{subfigure}
  \begin{subfigure}[t]{0.4\textwidth}
    \includegraphics[width=\textwidth]{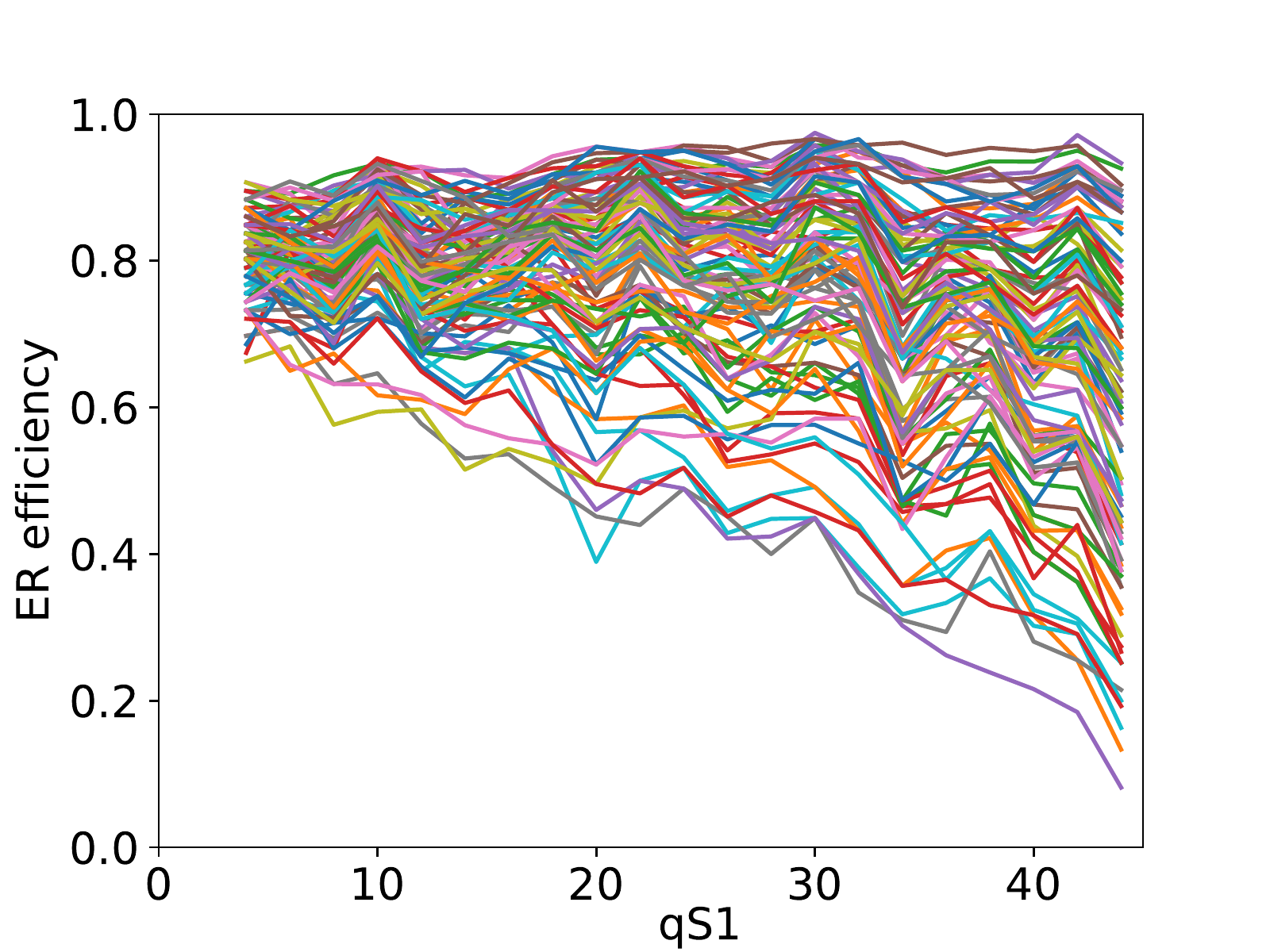}
    \caption{MLP 4Cr}
    \label{fig:4c_reg}
  \end{subfigure}
  \caption{Example of ER efficiencies as a function of $S1$ obtained
    from selected MLP model predictions. Each model is trained from
    scratch 100 times, thus 100 efficiency curves are obtained for
    each model. Fig.~\subref{fig:4c_reg} is obtained from model 4C
    with additional dropout layers.}
  \label{fig:er_eff}
\end{figure}

The BDT method has a unique attribute in that it demonstrates strong
differentiation capabilities for events below the NR median, while
maintaining a high level of efficiency for ER events that are dominant
above the NR median line in dark matter searches. It is crucial to
assess if the MLPs possess this characteristic.  Unfortunately, the
results indicate that the MLPs show inconsistent performance for ER
event recognition, with most models displaying low efficiency for ER
events, rendering the associated network parameters unsuitable for use
in data analysis. Examples of the ER efficiencies as function of $qS1$
for three given MLP models are presented in
Fig.~\ref{fig:er_eff}.

\begin{figure}[hbt]
  \centering
  \begin{subfigure}{.4\textwidth}
    \includegraphics[width=1.0\textwidth]{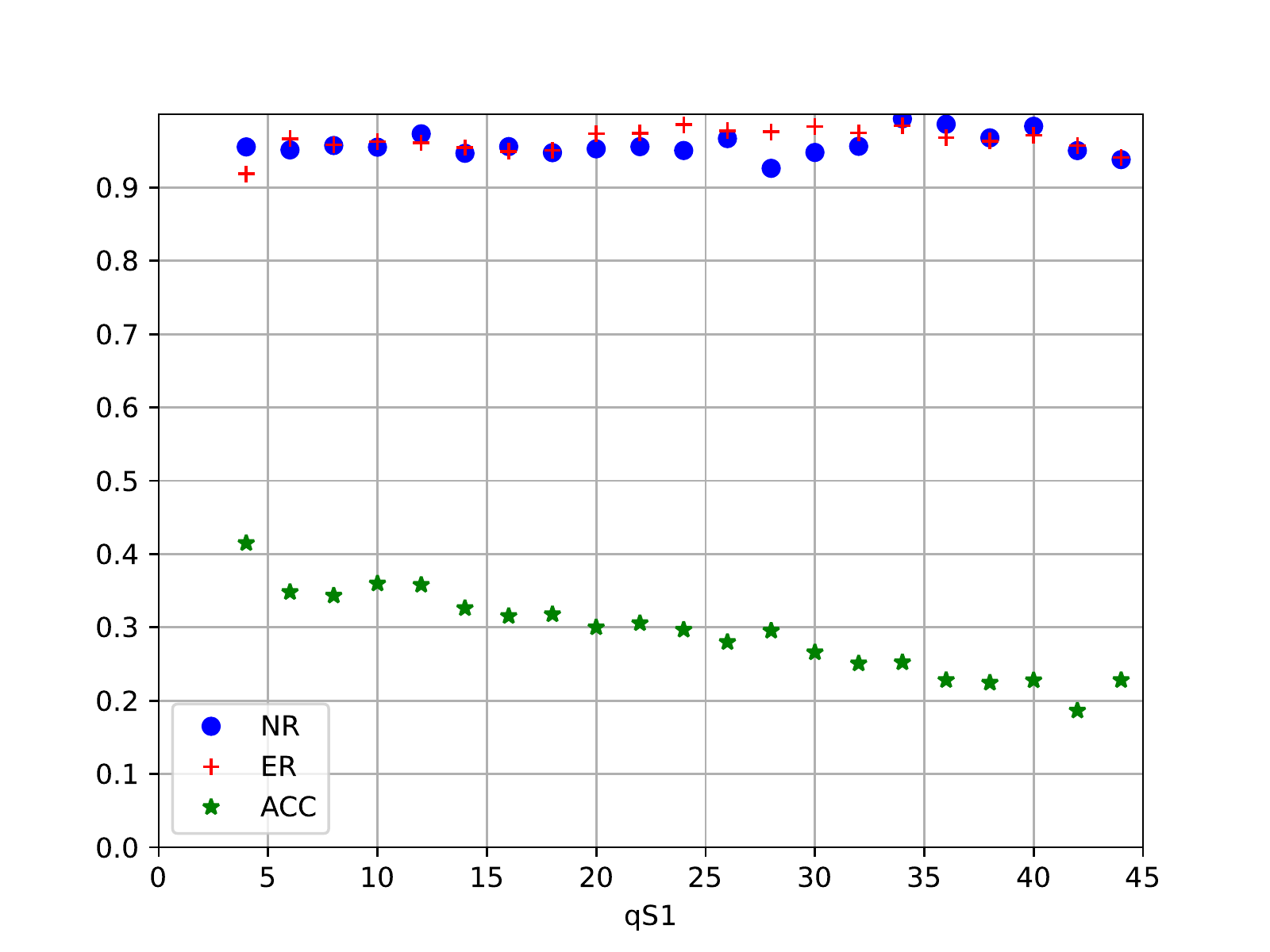}
    \caption{4C}
    \label{fig:best_4c}
  \end{subfigure}
  \begin{subfigure}{.4\textwidth}
    \includegraphics[width=1.0\textwidth]{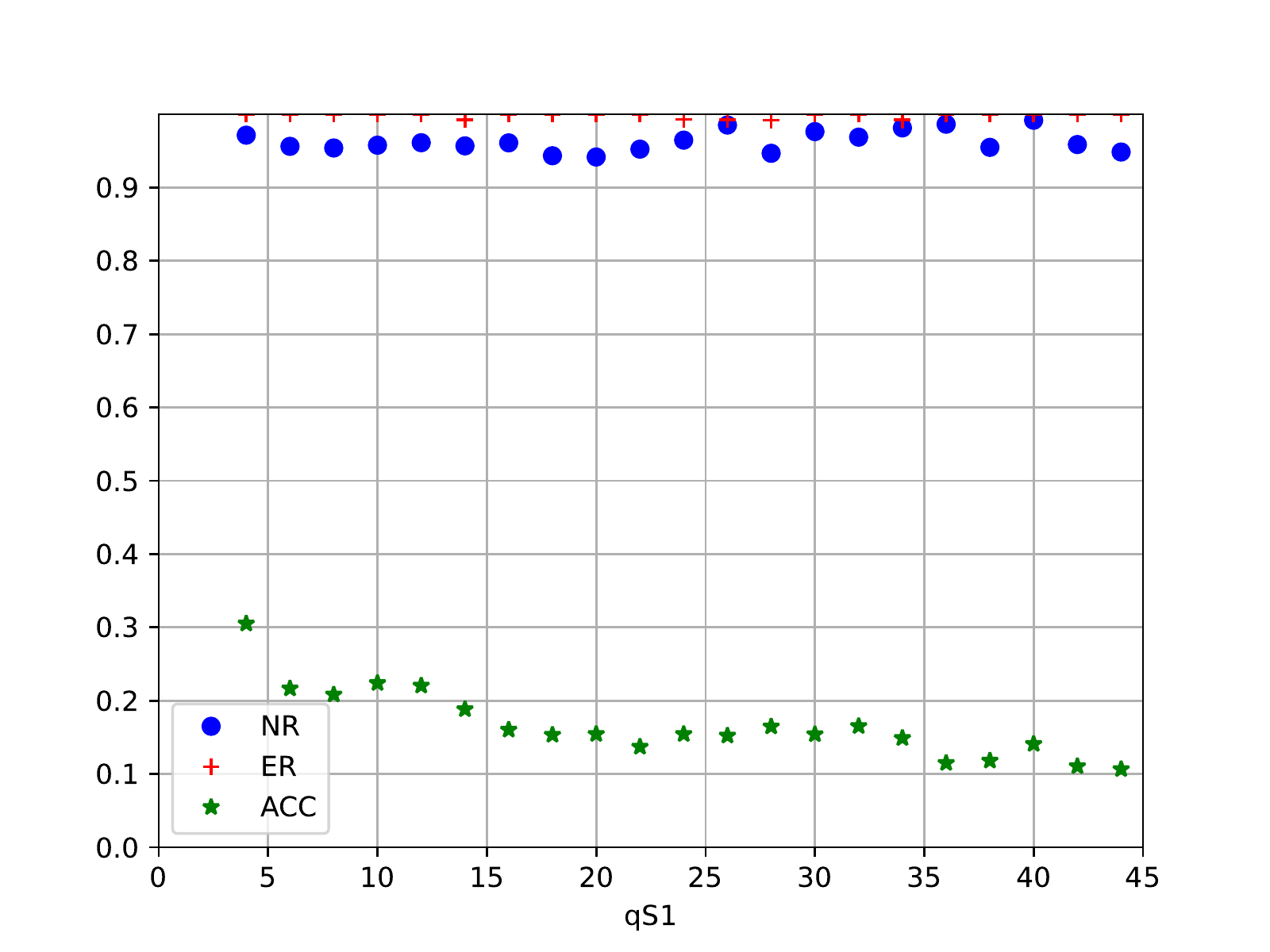}
    \caption{4Cu}
    \label{fig:best_4cu}
  \end{subfigure}
  \caption{The efficiencies obtained by the selected parameter set for MLP model 4C and 4Cu.}
  \label{fig:best_eff_4c_4cu}
\end{figure}

However, after a thorough examination of the results, some MLP
parameter sets were found to have both high discrimination power and
high efficiency for ER events.  One parameter set of model labeled
``4C'' is found to have the highest ER efficiency. The significance of
this model is 26.77. The efficiencies for the accidental background,
NR calibration events and ER calibration events are presented in
Fig.~\ref{fig:best_4c}.

The selected model are applied on the PandaX-II Run 11 dark matter
search data, resulting in 676 out of 708 candidate events having
survived\footnote{the post-unblinding cuts in Ref.~\cite{Wang:2020coa}
  is already applied}. The updated efficiencies and data were used to
drive the exclusion limit and sensitivity on the spin-independent
WIMP-nucleon cross section at a WIMP mass of 40~GeV/c$^2$. The results
are summarized in Table.~\ref{tab:limits}, alongside the results
obtained using the BDT method.  By using the selected 4C model, the
sensitivity is improved by 13\% in comparison with that obtained by
using the BDT method.

\begin{table}[hbt]
  \centering
  \begin{tabular}{ccccc}
    Method & Significance & Final candidates & Limit ($\times10^{-46}$cm$^2$) & Sensitivity ($\times10^{-46}$cm$^2$) \\\hline
    BDT & 26.2 & 693 & 2.51 & 2.39 \\
    MLP 4C & 26.77 & 676 & 5.68 & 2.07 \\
    MLP 4Cu & 27.48 & 698 & 3.64 & 1.87 \\
    MLP 4Eu & 27.64 & 697 & 4.15 & 1.92 \\\hline
  \end{tabular}
  \caption{The limits and sensitivities for WIMP with a mass of
    40~GeV/c$^2$, derived with different accidental suppressing
    methods applied, based on the PandaX-II Run 11 dark matter search
    data.}
  \label{tab:limits}
\end{table}

The application of regularization techniques was not successful in
curing the unstable behavior. Adding L2 regularization to each hidden
layer of Model 4C failed to discriminate between NR signal events and
the accidental background events. The fluctuation in the ER efficiency
curve remained even with the addition of dropout layers (dropout rate
0.2) after each hidden layer, as illustrated in Fig~\ref{fig:4c_reg}.

To enhance the stability of ER efficiencies in the MLP models, we
adopted a novel strategy for selecting the training data. We still
utilized the accidental background events below NR median, but
included all the NR events.  With the updated input data, we retrained
the 4C and 4E models, each 100 times, and renamed them to 4Cu and 4Eu,
respectively. All the trained models exhibit an ER efficiency close to
100\%, with high significance. The efficiencies obtained using the 4Cu
model with the highest significance is presented in
Fig.~\ref{fig:best_4cu}. The average significances are $26.82\pm0.17$
and $27.30\pm0.13$ for the 4Cu and 4Eu models, respectively. The
constraints and sensitivities from the models with highest
significances are presented in Table.~\ref{tab:limits}.  The limits
and sensitivities have both seen an improvement in comparison with the
old data selection strategy of DNN, suggesting that the revised
approach to preparing the input data for deep neural networks has led
to better results. Morever, model 4Cu shows a sensitivity improvement
of $21.8\%$ relative to the BDT method.

\section{Summary}
\label{sec:summary}
In this study, we explored the utilization of deep neural networks,
particularly MLPs, for suppressing accidental background in the
PandaX-II experiment. The results showed that increasing the number of
neurons in the layers improved the discrimination between signal and
background events. However, the traditional strategy of using only
data below the NR median during training resulted in unstable ER
efficiencies, making the DNNs challenging to use in data analysis. By
incorporating NR data above the NR median in the training, stable ER
efficiencies and high background suppression power were
achieved. Compared to the BDT method, the MLPs trained with the
updated data preparation strategy demonstrated improved sensitivity
for dark matter search.

In the field of dark matter direct detection, the application of deep
neural networks is not limited to signal and background
discrimination. For instance, generative adversarial networks can be
utilized to generate synthetic data that follows the same distribution
as actual data, which can be employed in both simulation and
analysis. We anticipate that machine learning techniques will continue
to provide even greater benefits in underground experiments.

\section*{Acknowledgment}
\label{sec:ack}
This project is supported in part by a grant from the Ministry of
Science and Technology of China (No. 2016YFA0400301), grants from
National Science Foundation of China (Nos. 12090063, 12175139), and by
Office of Science and Technology, Shanghai Municipal Government (grant
No. 18JC1410200). We also thank the sponsorship from the Chinese
Academy of Sciences Center for Excellence in Particle Physics
(CCEPP). We gratefully acknowledge the support of NVIDIA Corporation
with the donation of the Titan Xp GPU used for this research.

\bibliographystyle{JHEP} \bibliography{refs.bib}

\providecommand{\href}[2]{#2}\begingroup\raggedright\begin{thebibliography}{10}

\bibitem{Bertone:2004pz}
G.~Bertone, D.~Hooper and J.~Silk, \emph{{Particle dark matter: Evidence,
  candidates and constraints}},
  \href{https://doi.org/10.1016/j.physrep.2004.08.031}{\emph{Phys. Rept.}
  {\bfseries 405} (2005) 279}
  [\href{https://arxiv.org/abs/hep-ph/0404175}{{\ttfamily hep-ph/0404175}}].

\bibitem{Liu:2017drf}
J.~Liu, X.~Chen and X.~Ji, \emph{{Current status of direct dark matter
  detection experiments}},
  \href{https://doi.org/10.1038/nphys4039}{\emph{Nature Phys.} {\bfseries 13}
  (2017) 212} [\href{https://arxiv.org/abs/1709.00688}{{\ttfamily
  1709.00688}}].

\bibitem{Billard:2021uyg}
J.~Billard et~al., \emph{{Direct detection of dark matter\textemdash{}APPEC
  committee report*}},
  \href{https://doi.org/10.1088/1361-6633/ac5754}{\emph{Rept. Prog. Phys.}
  {\bfseries 85} (2022) 056201}
  [\href{https://arxiv.org/abs/2104.07634}{{\ttfamily 2104.07634}}].

\bibitem{PandaX:2016pdl}
{\scshape PandaX} collaboration, \emph{{Dark Matter Search Results from the
  Commissioning Run of PandaX-II}},
  \href{https://doi.org/10.1103/PhysRevD.93.122009}{\emph{Phys. Rev. D}
  {\bfseries 93} (2016) 122009}
  [\href{https://arxiv.org/abs/1602.06563}{{\ttfamily 1602.06563}}].

\bibitem{PandaX:2018wtu}
{\scshape PandaX} collaboration, \emph{{Dark matter direct search sensitivity
  of the PandaX-4T experiment}},
  \href{https://doi.org/10.1007/s11433-018-9259-0}{\emph{Sci. China Phys. Mech.
  Astron.} {\bfseries 62} (2019) 31011}
  [\href{https://arxiv.org/abs/1806.02229}{{\ttfamily 1806.02229}}].

\bibitem{Wu:2013cno}
Y.-C. Wu et~al., \emph{{Measurement of Cosmic Ray Flux in China JinPing
  underground Laboratory}},
  \href{https://doi.org/10.1088/1674-1137/37/8/086001}{\emph{Chin. Phys. C}
  {\bfseries 37} (2013) 086001}
  [\href{https://arxiv.org/abs/1305.0899}{{\ttfamily 1305.0899}}].

\bibitem{Li:2014rca}
J.~Li, X.~Ji, W.~Haxton and J.~S.~Y. Wang, \emph{{The second-phase development
  of the China JinPing underground Laboratory}},
  \href{https://doi.org/10.1016/j.phpro.2014.12.055}{\emph{Phys. Procedia}
  {\bfseries 61} (2015) 576} [\href{https://arxiv.org/abs/1404.2651}{{\ttfamily
  1404.2651}}].

\bibitem{Cheng:2017usi}
J.-P. Cheng et~al., \emph{{The China Jinping Underground Laboratory and its
  Early Science}},
  \href{https://doi.org/10.1146/annurev-nucl-102115-044842}{\emph{Ann. Rev.
  Nucl. Part. Sci.} {\bfseries 67} (2017) 231}
  [\href{https://arxiv.org/abs/1801.00587}{{\ttfamily 1801.00587}}].

\bibitem{Aprile:2009dv}
E.~Aprile and T.~Doke, \emph{{Liquid Xenon Detectors for Particle Physics and
  Astrophysics}}, \href{https://doi.org/10.1103/RevModPhys.82.2053}{\emph{Rev.
  Mod. Phys.} {\bfseries 82} (2010) 2053}
  [\href{https://arxiv.org/abs/0910.4956}{{\ttfamily 0910.4956}}].

\bibitem{PandaX-4T:2021bab}
{\scshape PandaX-4T} collaboration, \emph{{Dark Matter Search Results from the
  PandaX-4T Commissioning Run}},
  \href{https://doi.org/10.1103/PhysRevLett.127.261802}{\emph{Phys. Rev. Lett.}
  {\bfseries 127} (2021) 261802}
  [\href{https://arxiv.org/abs/2107.13438}{{\ttfamily 2107.13438}}].

\bibitem{Zhang:2016pgh}
T.~Zhang, C.~Fu, X.~Ji, J.~Liu, X.~Liu, X.~Wang et~al., \emph{{Low Background
  Stainless Steel for the Pressure Vessel in the PandaX-II Dark Matter
  Experiment}},
  \href{https://doi.org/10.1088/1748-0221/11/09/T09004}{\emph{JINST} {\bfseries
  11} (2016) T09004} [\href{https://arxiv.org/abs/1609.07515}{{\ttfamily
  1609.07515}}].

\bibitem{PandaX-4T:2021lbm}
{\scshape PandaX-4T} collaboration, \emph{{Low radioactive material screening
  and background control for the PandaX-4T experiment}},
  \href{https://doi.org/10.1007/JHEP06(2022)147}{\emph{JHEP} {\bfseries 06}
  (2022) 147} [\href{https://arxiv.org/abs/2112.02892}{{\ttfamily
  2112.02892}}].

\bibitem{Cui:2020bwf}
X.~Cui et~al., \emph{{Design and commissioning of the PandaX-4T cryogenic
  distillation system for krypton and radon removal}},
  \href{https://doi.org/10.1088/1748-0221/16/07/P07046}{\emph{JINST} {\bfseries
  16} (2021) P07046} [\href{https://arxiv.org/abs/2012.02436}{{\ttfamily
  2012.02436}}].

\bibitem{Wang:2022hkk}
Z.~Wang et~al., \emph{{Design and operation of the PandaX-4T high speed
  ultra-high purity xenon recuperation system}},
  \href{https://doi.org/10.1088/1748-0221/17/10/T10008}{\emph{JINST} {\bfseries
  17} (2022) T10008} [\href{https://arxiv.org/abs/2207.12632}{{\ttfamily
  2207.12632}}].

\bibitem{PandaX-II:2022waa}
{\scshape PandaX-II} collaboration, \emph{{Study of background from accidental
  coincidence signalsin the PandaX-II experiment}},
  \href{https://doi.org/10.1088/1674-1137/ac7cd8}{\emph{Chin. Phys. C}
  {\bfseries 46} (2022) 103001}
  [\href{https://arxiv.org/abs/2204.11175}{{\ttfamily 2204.11175}}].

\bibitem{Guest:2018yhq}
D.~Guest, K.~Cranmer and D.~Whiteson, \emph{{Deep Learning and its Application
  to LHC Physics}},
  \href{https://doi.org/10.1146/annurev-nucl-101917-021019}{\emph{Ann. Rev.
  Nucl. Part. Sci.} {\bfseries 68} (2018) 161}
  [\href{https://arxiv.org/abs/1806.11484}{{\ttfamily 1806.11484}}].

\bibitem{Carleo:2019ptp}
G.~Carleo, I.~Cirac, K.~Cranmer, L.~Daudet, M.~Schuld, N.~Tishby et~al.,
  \emph{{Machine learning and the physical sciences}},
  \href{https://doi.org/10.1103/RevModPhys.91.045002}{\emph{Rev. Mod. Phys.}
  {\bfseries 91} (2019) 045002}
  [\href{https://arxiv.org/abs/1903.10563}{{\ttfamily 1903.10563}}].

\bibitem{Schwartz:2021ftp}
M.~D. Schwartz, \emph{{Modern Machine Learning and Particle Physics}},
  \href{https://arxiv.org/abs/2103.12226}{{\ttfamily 2103.12226}}.

\bibitem{Qiao:2018edn}
H.~Qiao, C.~Lu, X.~Chen, K.~Han, X.~Ji and S.~Wang, \emph{{Signal-background
  discrimination with convolutional neural networks in the PandaX-III
  experiment using MC simulation}},
  \href{https://doi.org/10.1007/s11433-018-9233-5}{\emph{Sci. China Phys. Mech.
  Astron.} {\bfseries 61} (2018) 101007}
  [\href{https://arxiv.org/abs/1802.03489}{{\ttfamily 1802.03489}}].

\bibitem{NEXT:2020jmz}
{\scshape NEXT} collaboration, \emph{{Demonstration of background rejection
  using deep convolutional neural networks in the NEXT experiment}},
  \href{https://doi.org/10.1007/JHEP01(2021)189}{\emph{JHEP} {\bfseries 01}
  (2021) 189} [\href{https://arxiv.org/abs/2009.10783}{{\ttfamily
  2009.10783}}].

\bibitem{Khosa:2019qgp}
C.~K. Khosa, L.~Mars, J.~Richards and V.~Sanz, \emph{{Convolutional Neural
  Networks for Direct Detection of Dark Matter}},
  \href{https://doi.org/10.1088/1361-6471/ab8e94}{\emph{J. Phys. G} {\bfseries
  47} (2020) 095201} [\href{https://arxiv.org/abs/1911.09210}{{\ttfamily
  1911.09210}}].

\bibitem{EXO:2018bpx}
{\scshape EXO} collaboration, \emph{{Deep Neural Networks for Energy and
  Position Reconstruction in EXO-200}},
  \href{https://doi.org/10.1088/1748-0221/13/08/P08023}{\emph{JINST} {\bfseries
  13} (2018) P08023} [\href{https://arxiv.org/abs/1804.09641}{{\ttfamily
  1804.09641}}].

\bibitem{Domine:2019zhm}
{\scshape DeepLearnPhysics} collaboration, \emph{{Scalable deep convolutional
  neural networks for sparse, locally dense liquid argon time projection
  chamber data}},
  \href{https://doi.org/10.1103/PhysRevD.102.012005}{\emph{Phys. Rev. D}
  {\bfseries 102} (2020) 012005}
  [\href{https://arxiv.org/abs/1903.05663}{{\ttfamily 1903.05663}}].

\bibitem{LUX:2022vee}
{\scshape LUX} collaboration, \emph{{Fast and Flexible Analysis of Direct Dark
  Matter Search Data with Machine Learning}},
  \href{https://arxiv.org/abs/2201.05734}{{\ttfamily 2201.05734}}.

\bibitem{Wang:2020coa}
{\scshape PandaX-II} collaboration, \emph{{Results of dark matter search using
  the full PandaX-II exposure}},
  \href{https://doi.org/10.1088/1674-1137/abb658}{\emph{Chin. Phys. C}
  {\bfseries 44} (2020) 125001}
  [\href{https://arxiv.org/abs/2007.15469}{{\ttfamily 2007.15469}}].

\bibitem{Wu:2017cjl}
Q.~Wu et~al., \emph{{Update of the trigger system of the PandaX-II
  experiment}},
  \href{https://doi.org/10.1088/1748-0221/12/08/T08004}{\emph{JINST} {\bfseries
  12} (2017) T08004} [\href{https://arxiv.org/abs/1707.02134}{{\ttfamily
  1707.02134}}].

\bibitem{Brun:1997pa}
R.~Brun and F.~Rademakers, \emph{{ROOT: An object oriented data analysis
  framework}}, \href{https://doi.org/10.1016/S0168-9002(97)00048-X}{\emph{Nucl.
  Instrum. Meth. A} {\bfseries 389} (1997) 81}.

\bibitem{10.1162/neco.2006.18.7.1527}
G.~E. Hinton, S.~Osindero and Y.-W. Teh, \emph{A fast learning algorithm for
  deep belief nets},
  \href{https://doi.org/10.1162/neco.2006.18.7.1527}{\emph{Neural Comput.}
  {\bfseries 18} (2006) 1527–1554}.

\bibitem{tensorflow2015-whitepaper}
M.~Abadi, A.~Agarwal, P.~Barham, E.~Brevdo, Z.~Chen, C.~Citro et~al.,
  \emph{{TensorFlow}: Large-scale machine learning on heterogeneous systems},
  2015.

\bibitem{2014arXiv1412.6980K}
D.~P. {Kingma} and J.~{Ba}, \emph{{Adam: A Method for Stochastic
  Optimization}}, \href{https://doi.org/10.48550/arXiv.1412.6980}{\emph{3rd
  International Conference for Learning Representations} (2014) }
  [\href{https://arxiv.org/abs/1412.6980}{{\ttfamily 1412.6980}}].

\end{thebibliography}\endgroup
\end{document}